# Energy Efficient Clustering using Jumper Firefly Algorithm in Wireless Sensor Networks

Prof. N.V.S.N Sarma[#1], Mahesh Gopi[*2]

*Dept. of Electronics and Communication Engineering*

*National Institute of Technology Warangal, India*

*Abstract*— **Wireless Sensor Network (WSN) is a major and very interesting technology, which consists of small battery powered sensor nodes with limited power resources. The sensor nodes are inaccessible to the user once they are deployed. Replacing the battery is not possible every time. Hence in order to improve the lifetime of the network, energy efficiency of the network needs to be maximized by decreasing the energy consumption of all the sensor nodes and balancing energy consumption of every node. Several protocols have been proposed earlier to improve the network lifetime using optimization algorithms. Firefly is a metaheuristic approach. In this paper, Energy efficient clustering for wireless sensor networks using Firefly and Jumper Firefly algorithms are simulated. A new cost function has been defined to minimize the intra-cluster distance to optimize the energy consumption of the network. The performance is compared with the existing protocol LEACH (Low Energy Adaptive Clustering Hierarchy).**

*Keywords*— **Wireless Sensor Network, Clustering methods, Firefly algorithm, Jumper Firefly algorithm, Centralized algorithm, Energy efficient clustering.**

## I. INTRODUCTION

A Wireless Sensor Network is formed by spatially distributing low powered small sensor nodes communicating among themselves using radio signals and deployed randomly or manually in an unattended environment having limitations in power, sensing and processing capabilities. Due to recent developments in low powered micro sensor technologies the sensor nodes are available in large numbers at a low cost to be employed in a wide range of applications in environmental monitoring, military and many other areas [1]. In addition, the sensor nodes in a network are also capable of performing other functions such as data processing and routing. Wireless sensor nodes have certain constraints that make the designing wireless sensor network protocol difficult. One of the main constraints is limited power supply. So minimizing the energy consumption is a key requirement in the design of sensor network protocol and algorithms. In addition to this wireless sensor network design also demands other requirements such as fault tolerance, scalability, production costs, and reliability.

Since a large number of low-power sensor nodes have to be networked together, conventional techniques such as direct transmission from any specified node to the base station have to be avoided because in the direct transmission protocol, when a sensor node transmits data directly to the base station, the energy loss incurred can be quite extensive depending on the location of the sensor nodes relative to the base station. In such a scenario, the nodes that are farther away from the base station will dissipate the power heavily so that their batteries are drained much faster than those nodes that are closer to the base station [2]. A variety of improvements have been published for the last few years to limit the energy requirement in WSN, as mainly energy dissipation is more for wireless transmission and reception. Major approaches proposed so far are focused on making the changes at MAC layer and network layer. Two more major challenges are to fix the cluster heads over the grid and number of clusters in a network. To handle with all these challenges, clustering has been found to be the efficient technique.

*A. Clustering in Wireless Sensor Network*

Here, the sensor nodes are divided into small groups, which are called clusters as shown in the figure below. Each cluster will be having a cluster head (CH), which will monitor the remaining nodes. Nodes in a cluster do not communicate with the base station directly. They sense the data and send that to the cluster head. The cluster head will aggregate this, remove the redundant data and transmit it to the base station. So the energy consumption and number of messages transmitted to the base station will be reduced and number of active nodes in communication is also reduced. In this way the network lifetime is increased.

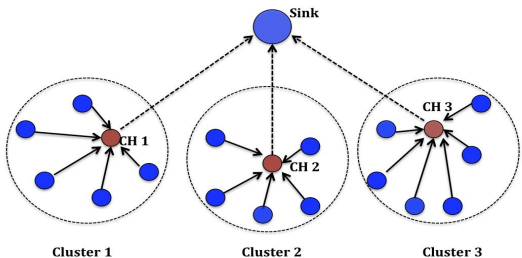

Fig 2 Clustering in Wireless Sensor Network

Generally clustering can be classified into three methodologies. First method is centralized clustering where the base station will configure the entire network into clusters, second method is distributed clustering where the sensor nodes configure themselves into clusters and third method is Hybrid clustering which is formed as the resulting configuration of the above two methods. Several protocols have been proposed earlier to maximize the sensor network lifetime. One of the well-known techniques is LEACH (Low Energy Adaptive Clustering Hierarchy) which is a distributed clustering algorithm [2]. Here nodes make decisions without any centralized control i.e. base station. It employs the technique of randomly rotating the role of a cluster head among all the sensor nodes in the network. All the nodes have a chance to become Cluster Head to balance the energy spent per round by each sensor node. Initially a node decides to be a Cluster Head with a probability p and broadcasts its decision. Specifically, after its election, an advertisement message is broad casted by each CH to the other sensor nodes and each one of the other (non- CH) nodes determines a cluster to belong to, by choosing the CH that can be reached using the least communication energy (based on the signal strength of each broadcast message). The application of meta-heuristic approaches like Particle Swarm Optimization (PSO) in Wireless Sensor Networks to solve the sensor network problems have been proposed earlier [3]. The main idea in the protocol is the selection of intra cluster distance between itself and the cluster member and optimization of energy management of the network. The main drawback of the PSO is its slow convergence in refined search space and weak local search ability.

In this paper, a centralized, energy aware cluster-based protocol to extend the sensor network lifetime by using Jumper Firefly algorithm has been developed. It makes use of a high-energy node as a cluster head and produces clusters that are evenly positioned throughout the whole sensor field. The main idea in the proposed protocol is the selection of intra-cluster distance between itself and the cluster member and optimization of energy management of the network.

The remaining part of the paper is organized as follows: Section II gives a brief description of the important papers that are reported. Section III introduces proposed system model for clustering the sensor network. Section IV discusses the implementation of the system for combined coverage and connectivity. Section V shows the experimental results achieved so far. Section VI presents conclusion and future work.

## II. RELATED WORK

Heinzelman et al studied the communication protocols, which can decrease the total energy dissipation of the WSNs [2]. They have proposed a new protocol called Low Energy Adaptive Clustering Hierarchy (LEACH), which is a distributed clustering approach. It randomly rotates the role of cluster heads among all the sensor nodes to evenly distribute the energy load among the sensors in the network and they compared the results with previous algorithms. The rotation is performed by getting each node to choose a random number T between 0 and 1. A node becomes a CH for the current round if the randomly chosen number is less than the following threshold:

$$T(n) = \begin{cases} \frac{p}{1-p\left(r\,mod\left(\frac{1}{p}\right)\right)} & , if\ n \in G \\ 0 & otherwise \end{cases} \quad (1)$$

Where, $p$ is the cluster head probability, $r$ is the current round number and $G$ is the group of nodes that have not been cluster-heads in the last $1/p$ rounds. After the CH selection, a TDMA schedule has been prepared by each cluster head and it transmits that schedule to all the cluster nodes in that respective cluster. With this the set up phase of LEACH will get completed.

W. B. Heinzelman et al proposed a centralized version of LEACH, LEACH-C, [4]. Unlike LEACH, where nodes self-configure themselves into clusters, in LEACH-C the base station will organize the network into clusters. In LEACH-C also there are two phases: setup phase and steady state phase. During the setup phase of LEACH-C, information regarding the location and energy level of each node in the network will be received by the base station. The base station finds a predetermined number of cluster heads and configures the network into clusters using the received information. The cluster groupings are chosen in such a way that the energy required for non-cluster head nodes to transmit their data to their respective cluster heads is minimum. Although the remaining operations of LEACH-C are identical to those of LEACH, results presented in [4] indicate a finite improvement over LEACH. The authors of [4] two main reasons for the improvement:

- The base station utilizes its global knowledge of the network to produce better clusters that require less energy for data transmission.
- The number of cluster heads in each round of LEACH-C equals a predetermined optimal value, whereas for LEACH the number of cluster heads varies from round to round due to the lack of global coordination among nodes.

N.A.Latiff, et al implemented [5] the energy aware clustering for wireless sensor networks using Particle Swarm Optimization (PSO) algorithm, which is a centralized algorithm. The main goal of PSO is to define the particle position that results in the best evaluation of a given cost function. During each generation, each particle uses the information about its earlier best individual position and global best position to update its candidate solution by utilizing the equations given below.

$$v_{id}(t) = w \times v_{id}(t-1) + c_1\varphi_1(p_{id} - x_{id}(t-1)) + c_2\varphi_2(p_{gd} - x_{id}(t-1)) \quad (2)$$

$$x_{id}(t) = x_{id}(t-1) + v_{id}(t) \quad (3)$$

Where $v$ is the particle velocity, $x$ is the particle position, $p_{id}$ is the best position of the particle, $p_{gd}$ is the global position and $w$ inertia weight. In WSN each particle represents the positions of the cluster heads.

V. Kumar et al proposed [6] an algorithm to maximize network lifetime in Wireless Sensor Networks (WSNs). The paths for data transfer are selected in such a way that the total energy consumed along the path is minimized. In order to support high scalability and fine data aggregation, frequently sensor nodes are grouped into non-overlapping; disjoint groups or subsets, which are called clusters. Clusters will create hierarchical Wireless Sensor Networks, which develop efficient utilization of limited resources (power) of sensor nodes, so that they extend lifetime of the network.

Wei Cheng el al [7] studied the impact of sensor nodes heterogeneity, in terms of their data amount and energy. They have proposed a novel distributed, adaptive, energy efficient clustering algorithm An adaptive Energy Efficient Clustering(AEEC) for wireless sensor networks, which is better suited for the heterogeneous wireless sensor networks. Here cluster heads will be selected based on the node energy relative to that of entire network, so that energy can be deployed more efficiently in heterogeneous networks. They assume that N nodes are distributed uniformly over an $M \times M$ square meters field A and the BS is located at the center of the field for simplicity. In this way the distance of any node to BS or its cluster head is less than $d_0$. They also considered that each node sends $l$ bit data per round, thereby the energy dissipated in the cluster head node during a round is given as:

$$E_{CH} = \left(\frac{N}{k} - 1\right) l . E_{elec} + \frac{N}{k} . l . E_{DA} + l . E_{elec} + l . \varepsilon_{fs} . d_{toBS}^2 \quad (4)$$

To the best of authors' knowledge, a lot of work has been done towards clustering, but none has studied the implementation of Jumper Firefly algorithm for the cluster formation.

III. SYSTEM MODEL

*A. Network Model*

In this section it is assumed that the network is composed of N different sensor nodes, which can sense, monitor and acquire information. They are randomly deployed uniformly within a M×M square region. The following postulates are also made for modeling the network.

1. All the nodes in the network are stationary and energy constrained.
2. Each node can perform sensing tasks periodically and always has the data to transfer to the Base Station.
3. Base station can be presented inside or outside the sensor network fields.
4. All nodes are capable of varying their transmitted power.
5. All nodes are eligible to operate either in cluster head mode or cluster member mode.
6. Data fusion is used to remove the redundant data so that the total data sent can be reduced.
7. Data communication is based on the single hop.

*B. Radio Energy Model*

The radio frequency energy consumption energy model for the sensor nodes is developed based on the first order radio model, as suggested in [7]. The first order radio model can be divided into free-space model and multi-path fading model depends on the distance between the transmitting node and receiving node. In that model the energy dissipation of transmitter is due to running the radio electronics and power amplifier. The dissipation of energy in receiver is because of radio electronics as shown in Figure 2. The energy dissipation is directly proportional to the square of the distance. For longer distances the energy consumption is proportional to $d^4$, where d is the distance between sender and receiver nodes. The energy consumption of $l$ bit message over a distance *d* between two nodes is given by

$$E_{TX}(l,d) = E_{TX-elec}(l,d) + E_{TX-amp}(l,d)$$

$$= \begin{cases} l.E_{elec} + l.\varepsilon_{fs}.d^2, & if \ d < d_0 \\ l.E_{elec} + l.\varepsilon_{mp}.d^4, & if \ d \geq d_0 \end{cases} \quad (5)$$

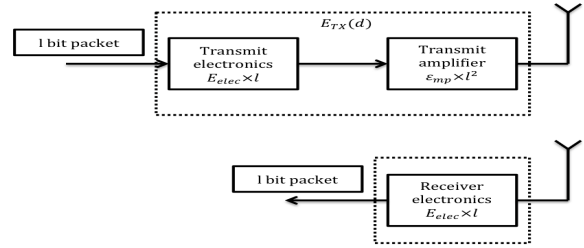

Fig 1 The First order radio model

While receiving an $l$ bit message, the radio spends:

$$E_{RX}(l,d) = l.E_{elec} \quad (6)$$

where $E_{TX}$ is radio energy dissipation for transmitter, $E_{TX}$ is the receiver radio energy dissipation, $E_{elec}$ is the energy consumption per bit to run the transmitter or the receiver circuit, $\varepsilon_{fs}$ and $\varepsilon_{mp}$ depend on the transmitter amplifier model used and $d_0$ is the threshold transmission distance which is given by

$$d_0 = \sqrt{\frac{\varepsilon_{fs}}{\varepsilon_{mp}}} \quad (7)$$

*C. Protocol Description*

*1. Firefly Algorithm (FFA)*

Firefly algorithm is one of the Meta-heuristic algorithms developed recently [6]. This is introduced by Dr. Xin She yang at Cambridge University in 2007, modeled after the flashing behavior of fireflies. The aim of Firefly Algorithm is to find the particle position that results in the best evaluation of a given fitness function. In this algorithm, there are three main rules:

- All fireflies are unisex i.e. a firefly will be attracted by other fireflies regardless of their sex.
- The attractiveness of a Firefly is directly proportional to its brightness and it decreases as the distance increases.
- The objective function results the brightness of a firefly.

The pseudo code for the Firefly algorithm can be prepared based on these three rules

*a) Attractiveness and Light Intensity* The light intensity varies according to the inverse square law i.e.:

$$I(r) = \frac{I_0}{r^2} \qquad (8)$$

Where I(r) is the light intensity at a distance $r$ and $I_0$ is the intensity at the source. When the medium is given, the light intensity can be determined as follows:

$$I(r) = I_0 \exp(-\gamma r) \qquad (9)$$

Where $\gamma$ is the absorption coefficient of the medium. In order to avoid the singularity at $r=0$ in eq(8), the following Gaussian form of the approximation is considered.

$$I(r) = I_0 \exp(-\gamma r^2) \qquad (10)$$

The attractiveness of a firefly is proportional to the light intensity seen by the adjacent fireflies. So the attractiveness $\beta$ of a firefly is given by the equation below.

$$\beta = \beta_0 \exp(-\gamma r^m) \qquad (11)$$

Where $\beta_0$ is the attractiveness at $r=0$.

*Distance* $r_{i,j}$ is the distance between any two fireflies $i$ and $j$ which are located at $x_i$ and $x_j$ respectively, the Cartesian distance is given by the equation below

$$r_{i,j} = \sqrt{\sum_{k=1}^{d}(x_{i,k} - x_{j,k})^2} \qquad (12)$$

Where $x_{i,k}$ is the $k^{th}$ component of the spatial coordinate $x_i$ of the $i^{th}$ firefly and $d$ is the number of dimensions.

C. Movement

The movement of a firefly $i$ towards more attractive (brighter) firefly $j$ is given by

$$x_i = x_i + \beta_0 e^{-\gamma r_{i,j}^2}(x_j - x_i) + \alpha \epsilon \qquad (13)$$

Where the second term is due to the attraction while the third term is randomization via $\alpha$ being the randomization parameter.

*d) Cluster setup using Firefly algorithm:* The protocol is a centralized algorithm in which all the clustering procedures will be done at the base station. The Base Station runs Firefly Algorithm to determine the best $K$ CHs that can minimize the cost function, which is defined in [5] as

$$cost = \beta \times f_1 + (1 - \beta) \times f_2 \qquad (14)$$

$$f_1 = max_{k=1,2,...K}\{\sum_{\forall n_i \in C_{p,k}} d(n_i, CH_{p,k}) / |C_{p,k}|\} \qquad (15)$$

$$f_2 = \frac{\sum_{i=1}^{N} E(n_i)}{\sum_{k=1}^{K} E(CH_{p,k})} \qquad (16)$$

Where $f_1$ is the maximum average Euclidean distance of nodes to their associated cluster heads and $|C_{p,k}|$ is the number of nodes that belong to cluster $C_k$ of particle p. $f_2$ is the function which is the ratio of total initial energy of all the nodes ($n_i$, i = 1, 2…. N), in the network to the total current energy of the cluster head candidates in the current round. The constant value $\beta$ is a user defined constant to weigh the contribution of each of the sub-objectives. For a sensor network with $N$ nodes and $K$ predetermined number of clusters, the wireless sensor network can be clustered as follows:

1. Initialize S particles to contain K randomly selected cluster heads among the eligible cluster head candidates.

2. Calculate the cost function of each particle:

   i. For each node $n_i$, i=1,2,....N.

   a) Calculate distance $d(n_i, CH_{p,k})$ between node $n_i$ and all cluster heads $CH_{p,k}$.

   b) Assign node $ni$ to cluster head $CH_{p,k}$ where: $d(n_i, CH_{p,k}) = min\{d(n_i, CH_{p,k})\}$ for $k = 1,2 … K$ (17)

  ii. Calculate the cost function using equations (14) to (16).

3. Rank the fireflies and find the current best.
4. Update the position of the particle.
5. Limit the change in the particle's position value.
6. Map the new updated position with the closest (x, y) coordinates.
7. Repeat the steps 2 to 6 until the maximum number of iterations is reached.

The base station has identified the optimal set of cluster heads and their associated cluster members. The base station transmits information that contains the cluster head ID for each node back to all nodes in the network.

*e) Advantages and disadvantages of Firefly algorithm:* Firefly algorithm seems to be a favorable optimization tool phenomenon due to the effect of the attractiveness function, which is a unique to the firefly behavior. Firefly not only includes the self-improving process with the current space, but it also includes the improvement among its own space from the previous stages. Firefly algorithm has some disadvantages such as getting trapped into several local optima. Firefly algorithm sometimes performs local search as well and sometimes is unable to completely get rid of them. Firefly algorithm parameters are fixed and they do not change with the time. In addition Firefly does not memorize or remember any history of better situation, and they may end up missing their situations.

**Algorithm1:** *Clustering using Firefly Algorithm*

```
Data:
    • Generate S particles to contain K randomly selected cluster heads.
    • Map the randomly generated positions with the closest (x,y) coordinates.
Result: Cluster heads positions are obtained
while(t< Max Generation)
Map the positions with the closest (x,y) coordinates
Evaluate the cost function
for i=1 to n (all n fireflies)
    for j=1 to n (all n fireflies)
        if (I_j > I_i)
            Update the particles positions;
            Limit the change in the particle's position value;
        end
    end
end
Rank the fire flies and find the current best;
end
Post process the results;
End procedure;
```

*2) Jumper Firefly Algorithm (JFA):*
There is diversity among the members in every population of live creatures in terms of quality and fitness. Normally Members with high fitness will perform their jobs efficiently and reach high quality achievements. However the low quality members cannot reach such high quality achievements. The quality of each population is evaluated by individual members of the respective population. The probability of obtaining eligible and suitable solutions in qualified populations is high. So Mahdi Bidar and HamidrezaRashidyKanan developed a new algorithm [9] based on firefly algorithm that to improve the performance of the agents in determining more appropriate solutions by modifying them (by making change in the agents situations), develops the quality of firefly's society, thereby the probability of finding the optimal solution can be increased. For this process, they have used a Status Table, which records and observes all the details of the Fireflies behavior. This status table helps to indicate the agents that have to be changed in their situations by jumping into new situations. The process of decision-making is executed on the search agents to realize whether any agent needs to use the jump option or not. The Status Table used in this algorithm is shown below.

TABLE I

Parameters Summary

| Firefly | 1 | 2 | 3 | ... | n |
|---|---|---|---|---|---|
| Parameter | | | | | |
| Situation | | | | | |
| Fitness | | | | | |
| Worst | | | | | |
| Qualification | | | | | |

In the Status Table, Situation defines the location of each and every firefly in search space at $i^{th}$ stage; Fitness defines the quality of solution found by the fireflies at $i^{th}$ stage; The number of worst solutions (in each step in comparison with the other agent's) attained by every firefly during searching phases is indicated by Worst. Finally, Qualification indicates the cost of each firefly from starting of the search process to the investigation moment of the table. In this table, Qualification at $i^{th}$ phase gets update as

$$Qualification(i) = fitness(i) + Qualification(i-1) \quad (18)$$

Based on Status Table, firefly (m) is in hazard and needs to use jump option if its values in Status table is as shown below

$$worst(firefly(m)) = \max(worst) > \eta \quad (19)$$

$$Qualification(firefly(m)) = \min(Qualification) \quad (20)$$

$$Qualification(firefly(m)) < \text{AVE}(Qualification) - \Omega \quad (21)$$

AVE(Qualification) indicates the mean of the Qualification row in Status Table, $\eta$ is a user defined variable which will prevent a firefly from jumping in the beginning of the searching process in the new algorithm and Ω is a constant. Based on hazard condition, agents involved in hazard are the agents that have found low quality solutions and there is no hope to change the searching trend in finding proper solutions. So, the algorithm provides the agents with jump option by which they can rearrange themselves in a new situation and start a new life. New situation for fireflies means the re-initialization and rearranging in new location. Since the re-initialization must preferably be done randomly, the algorithm places the weakest firefly stochastically in the new location. Update of Status Table for the agents using the jump option is as shown below.

$$Situation(firefly(m)) = new\ situation\ of\ firefly(m) \quad (22)$$

$$fitness(firefly(m)) = fitness\ of\ newly\ found\ solution \quad (23)$$

$$Qualification(firefly(m)) = \text{AVE}(Qualification) \quad (24)$$

Now our aim is to implement this Jumper Firefly algorithm in WSN for clustering.

a) *Cluster setup using Jumper Firefly algorithm*
   Cluster setup using Jumper Firefly algorithm is similar to that using Firefly algorithm but Jumper Firefly algorithm is implemented at the base station instead of Firefly algorithm. The new algorithm is as shown below

**Algorithm2:** *Clustering using Jumper Firefly Algorithm*

Data:
- Generate $S$ particles to contain $K$ randomly selected cluster heads.
- Define the user defined constants like $\eta$, $\Omega$, $\beta$...
- Map the randomly generated positions with the closest (x,y) coordinates.
- Create the Status table

Result: Cluster heads positions are obtained

**while**(t< *Max Generation*)
**if**(*any firefly is in hazard*)
**Put the firefly in new position stochastically;**
**Update the Status table;**
**end**
Map the positions with the closest (x,y) coordinates
Evaluate the cost function
**for** i=1 to n (all n fireflies)
**for** j=1 to n (all n fireflies)
**if**($I_j > I_i$)
Update the particles positions;
Limit the change in the particle's position value;
**Update the Status table;**
**end**
**end**
**end**
Rank the fire flies and find the current best;
**end**
Post process the results;
End procedure;

## IV. SIMULATION RESULTS

The performance of the Firefly algorithm is evaluated using MATLAB. The simulated network is for 100 nodes in a 200m×200m network area and the base station is located at the center of the area. The following table gives the data considered during the simulation of the network. The performance of new protocol is compared with the Firefly algorithm (FFA) and LEACH. A snapshot of uniformly distributed sensor network is as shown in figure 3. The number of clusters is set to be 5 percent of the total number of sensor nodes. The simulations continued until all the nodes consumed all their energy.

TABLE II

Parameters Summary

| Parameter | Value |
|---|---|
| $E_0$ | 0.2J |
| $E_{elec}$ | 50nJ/bit |
| $E_{DA}$ | 50nJ/bit/signal |
| $\varepsilon_{fs}$ | 10pJ/bit/$m^2$ |
| $\varepsilon_{mp}$ | 0.0013pJ/bit/$m^2$ |
| $l$ | 4000 bits |

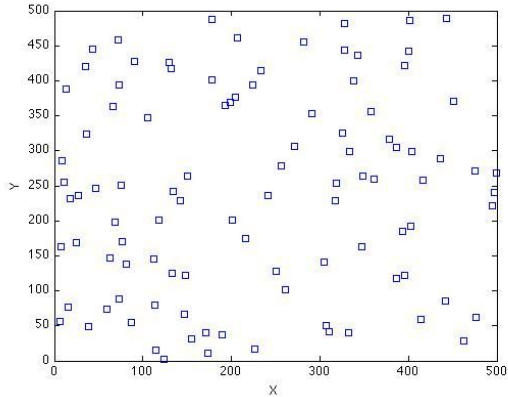

Fig 3  A snapshot of uniformly distributed Sensor Network

Figure 4 shows the network lifetime, which clearly demonstrates that the proposed protocol can prolong the network life time when compared with LEACH and Firefly algorithm protocols. This is because Jumper Firefly algorithm gives better network partitioning with minimum intra-cluster distance and also cluster heads that are optimally distributed across the network. So the energy consumed by all the nodes for communication can be reduced drastically because the distance between cluster members and their cluster heads are less.

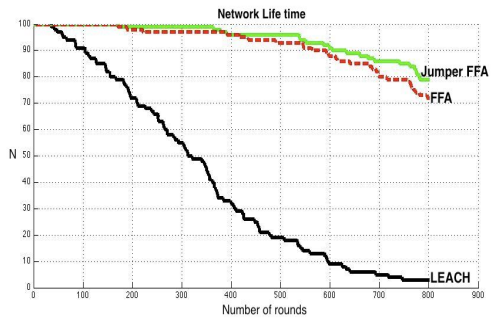

Fig 4  Network lifetime for the nodes with heterogeneous energy

### V. CONCLUSION

In this paper an energy aware cluster based protocol for wireless sensor networks using Jumper Firefly algorithm is implemented for the first time. A new cost function is mentioned, which can consider the maximum distance between the cluster head and cluster members and the remaining energy of the cluster head candidates in the optimization procedure.

Results from the simulations indicate that clustering using Jumper Firefly algorithm gives a better network lifetime when compared to Firefly and LEACH. Future scope includes the implementation of hybrid optimization technique for clustering in wireless sensor networks.